\documentclass[lettersize,journal]{IEEEtran}
\usepackage{amsmath,amsfonts}
\usepackage{algorithmic}
\usepackage{algorithm}
\usepackage{array}
\usepackage[caption=false,font=normalsize,labelfont=sf,textfont=sf]{subfig}
\usepackage{textcomp}
\usepackage{stfloats}
\usepackage{url}
\usepackage{verbatim}
\usepackage{graphicx}
\usepackage{cite}
\usepackage{booktabs}
\usepackage{array}
\usepackage{adjustbox}
\usepackage{multirow}
\usepackage{makecell}
\usepackage{xurl}
\usepackage[T1]{fontenc}

\usepackage{longtable}

\begin{document}

\title{Biological Sequence Clustering: A Survey}

\author{Simeng Zhang, Xinying Liu, Jun Lou, Mudi Jiang, Quan Zou, Zengyou He
\thanks{S. Zhang, X. Liu, J. Lou, M. Jiang, Z. He are with School of Software,
Dalian University of Technology, Dalian, China. Q. Zou is with Institute of Fundamental and Frontier Sciences, University of Electronic Science and Technology of China, Chengdu, China\\E-mail: zyhe@dlut.edu.cn}
}


\maketitle

\begin{abstract}
The rapid development of high-throughput sequencing technologies has led to an explosive increase in biological sequence data, making sequence clustering a fundamental task in large-scale bioinformatics analyses. Unlike traditional clustering problems, biological sequence clustering faces unique challenges due to the lack of direct similarity measures, strict biological constraints, and demanding requirements for both scalability and accuracy. Over the past decades, a wide variety of methods have been developed, differing in how they model sequence similarity, construct clusters, and prioritize optimization objectives.
In this review, we provide a comprehensive methodological overview of biological sequence clustering algorithms. We begin by summarizing the main strategies for modeling sequence similarity, which can be divided into three stages: sequence encoding, feature generation, and similarity measurement. Next, we discuss the major clustering paradigms, including greedy incremental, hierarchical, graph-based, model-based, partitional, and deep learning approaches, highlighting their methodological characteristics and practical trade-offs. We then discuss clustering objectives from three key perspectives: scalability and resource efficiency, biological interpretability, and robustness and clustering quality. Organizing existing methods along these dimensions allows us to explore the trade-offs in biological sequence clustering and clarify the contexts in which different approaches are most appropriate. Finally, we identify current limitations and challenges, providing guidance for researchers and directions for future method development.

\end{abstract}


\section{Introduction}
\IEEEPARstart{T}{he} rapid evolution of high-throughput sequencing technologies has driven an exponential increase in the scale of biological sequence data over the past decade \cite{saw2019alignment} \cite{katz2022sequence}. For such massive datasets, effective representation, compression, and data analysis have become both critical and challenging tasks in bioinformatics. In this context, biological sequence clustering has emerged as a key strategy, playing an indispensable role in microbial community analysis \cite{sun2012large}, protein family construction \cite{gu2002diverge}, virus evolution\cite{bustamam2017application}, and database redundancy removal \cite{chen2017duplicates} \cite{sikic2010protein}. 
Particularly since 2021, a large number of methods \cite{melnyk2020clustering,ali2021effective,chen2024clustering,tayebi2021robust,chourasia2021clustering} have been proposed for clustering SARS-CoV-2 variants to analyze differences and relationships among viral strains and to infer their potential evolutionary trajectories, highlighting the substantial practical value of biological sequence clustering.
It is essential for revealing structure and patterns within large scale sequence collections and encompasses the entire analytical workflow, from raw data preprocessing to biological interpretation, thereby constituting a foundational component of modern bioinformatics.

As data volume and diversity continue to increase rapidly, sequence clustering now faces substantial challenges. The dramatic growth in sequence numbers renders methods based on pairwise alignment or global distance matrices computationally inefficient for large-scale datasets \cite{luczak2019survey}. At the same time, similarity patterns vary markedly across different types of biological sequences. Short reads, complete genomes, and distantly homologous proteins differ significantly in their distributional characteristics, evolutionary divergence, and noise structure, making it difficult to apply a universal clustering threshold across datasets. In addition, similarity computation methods are highly diverse, including identity measures derived from alignments, distances based on $k$-mers  \cite{li2006cd} or sketches \cite{mash}, embedding space distances \cite{embedding}, and graph-based weights \cite{Mpick}. Furthermore, the emergence of deep learning and high dimensional embedding techniques has introduced new challenges related to the interpretability of clustering results. Collectively, considerations of time complexity, memory consumption, parameter sensitivity, and robustness have become critical constraints, making the development of scalable, accurate, and interpretable clustering models a central focus of current research.

In response to the rapid increase in tools and methods in recent years, several research efforts \cite{li2012ultrafast,wzgreview, zoureview,methodsreview,matar2025biological,Chenreview,ju2025comparative,ju2024depth,chen2013comparison,russell2010grammar} have compared existing sequence clustering approaches through systematic experimental validations, thereby providing bioinformatics researchers with clear and practical guidance for proper method selection.
However, these existing review-type articles on biological sequence clustering still have notable limitations. First, most reviews focus primarily on empirical performance comparisons and lack a systematic categorization and analysis of different methods from the perspective of algorithmic design. Second, they provide insufficient coverage of emerging clustering paradigms, therefore fail to capture recent advances in the field. Consequently, these existing reviews lack a methodological decomposition and synthesis of the clustering workflow and pay limited attention to the mathematical principles, and the key issues that algorithmic designs are intended to address.

To fill this gap, this study provides a systematic categorization of existing biological sequence clustering methods starting from their algorithmic structure. Our discussion is organized around three critical dimensions that characterize the clustering model, technical roadmap, and application scenario of different approaches:

\begin{enumerate}
	\item Similarity computation methods, including strategies based on alignments, $k$-mers, sketches, and graphs;
	\item Clustering principles and workflows, encompassing hierarchical clustering, greedy incremental clustering, graph-based clustering, partitional clustering, deep learning-based clustering and model-driven paradigms;
	\item Design objectives and application scenarios, distinguishing methods by their optimization priorities in scalability, memory efficiency, interpretability, and robustness.
\end{enumerate}

This work aims to provide researchers with a clear algorithm oriented classification framework that facilitates method selection, refinement, and innovation, and offers theoretical support and practical guidance for clustering analysis of large-scale and highly diverse biological sequence datasets.

The remainder of this paper is organized as follows. Section I of this paper discusses the background of biological sequence clustering and introduces the structure of the article. Section II highlights the differences between biological sequence clustering and general data clustering, and the challenges involved. Section III presents a three-layer methodological framework that decomposes and analyzes the clustering process. Section IV outlines three main steps in biological sequence similarity modeling. Section V summarizes and categorizes the typical methods of existing biological sequence clustering mechanisms. Section VI discusses the three main objectives of biological sequence clustering algorithms and the approaches to address them. Finally, Section VII provides a summary of the current development and the main challenges in biological sequence clustering.

\section{Challenges for biological sequence clustering}
Clustering, as an unsupervised learning task, aims to group data so that samples within the same cluster are similar, while samples from different clusters are dissimilar. In standard vector data clustering, objects are typically represented as fixed dimensional numerical vectors, and similarity measures have clear geometric interpretations, such as Euclidean distance or cosine similarity. However, when dealing with biological sequences, a series of unique challenges arise that make general clustering methods difficult to apply directly.

Biological sequences are variable-length strings of discrete symbols, such as DNA, RNA, or protein sequences, composed of characters from a limited alphabet. They often differ greatly in length, and operations such as insertions, deletions, and substitutions have clear biological meaning. As a result, there is no natural one to one correspondence between positions in different sequences, for example, a series of gaps between two sequences of different lengths may represent several separate mismatches or a single evolutionary insertion or deletion event. Measuring such differences accurately depends on assumptions about the evolutionary process, rather than on simple counts of character mismatches. As a result, traditional distance measures based on vector representations and position by position comparisons are often unsuitable for biological sequences. Measures such as alignment identity or alignment score have therefore been widely regarded as standard indicators of biological sequence similarity. However, despite their strong biological meaning, alignment-based methods are computationally costly and become impractical for large-scale biological sequences. 

With the advent of high throughput sequencing technologies, although pairwise sequence alignment provides extremely accurate similarity values, its computational cost has driven researchers toward similarity modeling approaches that do not rely on explicit alignments \cite{zielezinski2017alignment,ondov2016mash,kucherov2019evolution}. These methods encode sequences, capture local patterns, or estimate information structure, trading some degree of biological precision for substantial gains in computational efficiency.

Beyond the challenges associated with defining and computing similarity, biological sequence clustering is also subject to distinctive constraints at the level of clustering mechanisms.
Traditional clustering methods often aim to optimize a global objective and assume that clusters have regular shapes or simple distributions. In contrast, biological sequence data typically show highly uneven similarity patterns. Many sequences may be very similar or even redundant, while a small portion of sequences are highly diverse, leading to cluster structures with clear hierarchical and long tail patterns. These features are closely related to evolutionary history and species diversity. Applying clustering methods that assume simple geometric structures or global distribution models can therefore be both computationally expensive and biologically unrealistic.

Taken together, biological sequence clustering differs fundamentally from standard clustering problems in terms of data structure, the semantics of similarity, computational scale, and structural assumptions. 
For these reasons, biological sequence clustering has long relied on specialized tools, such as alignment-based similarity measures and fixed similarity thresholds, and has gradually developed its own set of methods. From a methodological viewpoint, the differences among existing approaches can be understood along three main dimensions.
First, how sequence similarity is modeled while balancing biological meaning and computational feasibility. Second, given a definition of similarity, which clustering mechanisms are used to construct cluster structures. Third, the design objectives of different methods with respect to scalability, accuracy, memory efficiency, and robustness. The following sections systematically review existing approaches along these three dimensions and clarify their core principles and application scenarios.

\section{A Taxonomy of Biological Sequence Clustering}

\begin{figure*}[!ht] 
	\centering 
	\includegraphics[width=\textwidth]{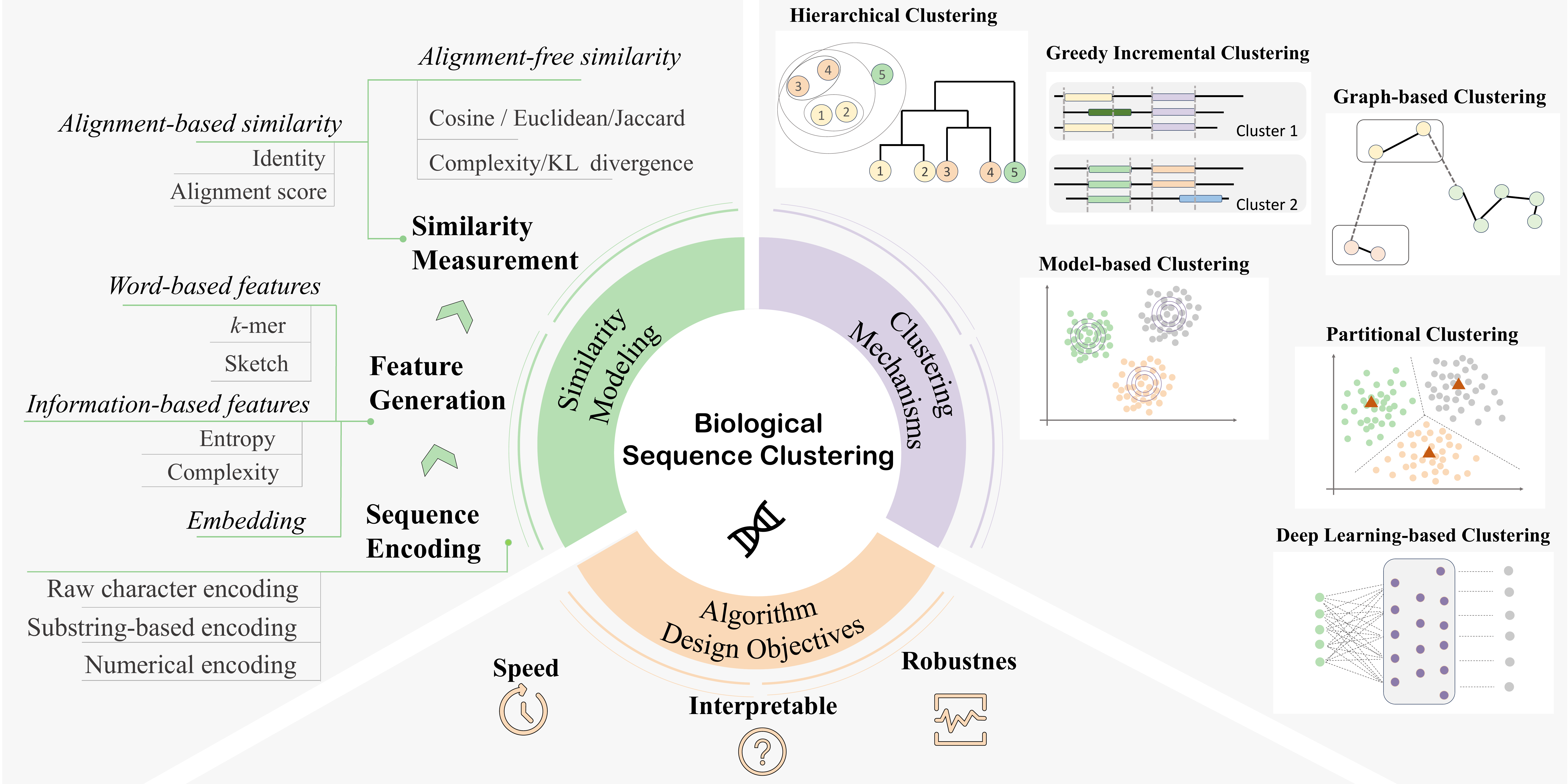} 
	\caption{A methodological overview of biological sequence clustering. } 
	\label{overview} 
\end{figure*}

Having clarified the unique challenges inherent to biological sequence clustering, a natural question arises: how do existing methods address these issues? To systematically organize current approaches and reveal their underlying connections, as shown in Fig.~\ref{overview}, this paper introduces a three layer methodological framework that decomposes and analyzes the clustering process.

The central idea of this framework is to view biological sequence clustering as an integrated workflow composed of \textbf{similarity modeling}, the choice of \textbf{clustering mechanism}, and \textbf{algorithm design objectives}. The fundamental differences among methods often lie not in implementation details, but in the distinct trade offs they make across these three levels. By adopting this framework, diverse approaches can be compared from a unified perspective, enabling a clearer understanding of their design motivations and appropriate application contexts.

\subsection{Similarity Modeling: How to Define and Compute Relationships Between Sequences?}

Similarity modeling constitutes the starting point of biological sequence clustering algorithms. As discussed above, similarity between sequences is not inherently defined but rather depends on a balance between biological semantics and computational feasibility. Consequently, all methods must first address a fundamental question: how should the similarity between two sequences be defined, and how can this similarity measure be calculated efficiently. 

At this level, methodological differences are primarily reflected in how sequences are abstracted and encoded \cite{alignfree2023}. One class of approaches characterizes character level correspondence through sequence alignment, emphasizing biological interpretability. Another class relies on sequence encoding and feature extraction to approximate global or local similarity structures without explicit alignment, thereby achieving greater scalability. In many cases, similarity information is further organized into matrix or graph representations to provide more suitable inputs for subsequent clustering steps. Design choices made at the similarity modeling stage directly determine the biological meaning of clustering results and the scale of data to which a method can be effectively applied.

\subsection{Clustering Mechanisms: How to Obtain Clusters Based on Sequence Similarity?}

Once similarity or distance relationships between sequences have been established, the clustering mechanism is responsible for constructing cluster structures based on this information. Compared with general purpose clustering, biological sequence clustering exhibits clear preferences on how to design the clustering procedure, reflecting a combined consideration of data scale, similarity distribution characteristics, and requirements for biological interpretability.

In this study, existing methods are systematically summarized into seven categories: greedy incremental clustering, hierarchical clustering, graph-based clustering, partitional clustering, model-driven clustering and deep learning-based clustering. For example, greedy incremental methods achieve efficient clustering by comparing sequences to representative centroids using predefined thresholds and are well suited for large scale datasets. Hierarchical clustering relies on pairwise similarities to iteratively merge or split clusters, yielding results with good interpretability. Graph-based approaches represent sequences as nodes and similarities as edges, and then partitions graph into subgraphs to obtain clusters. Model-driven methods apply model-based clustering methods such as Gaussian mixture model to embedded representations of sequences. 

\subsection{Algorithm Design Objectives: What Core Problems Do Methods Aim to Address?}

Beyond similarity modeling and clustering mechanisms, another important source of variation among biological sequence clustering methods lies in their algorithm design objectives. Unlike standard clustering, which often focuses primarily on clustering quality, biological sequence clustering typically requires balancing multiple, and sometimes competing, goals.

In large scale data analysis, scalability and memory requirements become primary concerns, driving the development of approximate similarity computation and efficient clustering strategies. In datasets with substantial noise, robustness is particularly critical. In contrast, for tasks such as phylogenetic analysis or functional annotation, the biological interpretability and consistency of clustering results are often more important than computitional efficiency alone. These design objectives permeate the choices made in both similarity modeling and clustering mechanisms and together shape the overall characteristics of different methods.

Together, these three dimensions form a comprehensive categorization framework.  Taking RabbitTClust \cite{rabbit} as an example, it adopts $k$-mer based sketch representations for similarity computation, employs greedy or minimum spanning tree based strategies at the clustering mechanism level, and is explicitly optimized for efficient processing of ultra large scale genomic datasets. In contrast, methods such as SpCLUST \cite{SpCLUST} and GCLUST \cite{GCLUST} enhance their ability to capture complex clustering structures through spectral embedding and model-based clustering. By situating algorithms within this three-layer framework, it becomes possible to more clearly understand their design objectives, application scenarios, and roles in a broader landscape of sequence clustering techniques.

\section{Similarity Modeling}
\begin{figure*}[!ht] 
	\centering 
	\includegraphics[width=\textwidth]{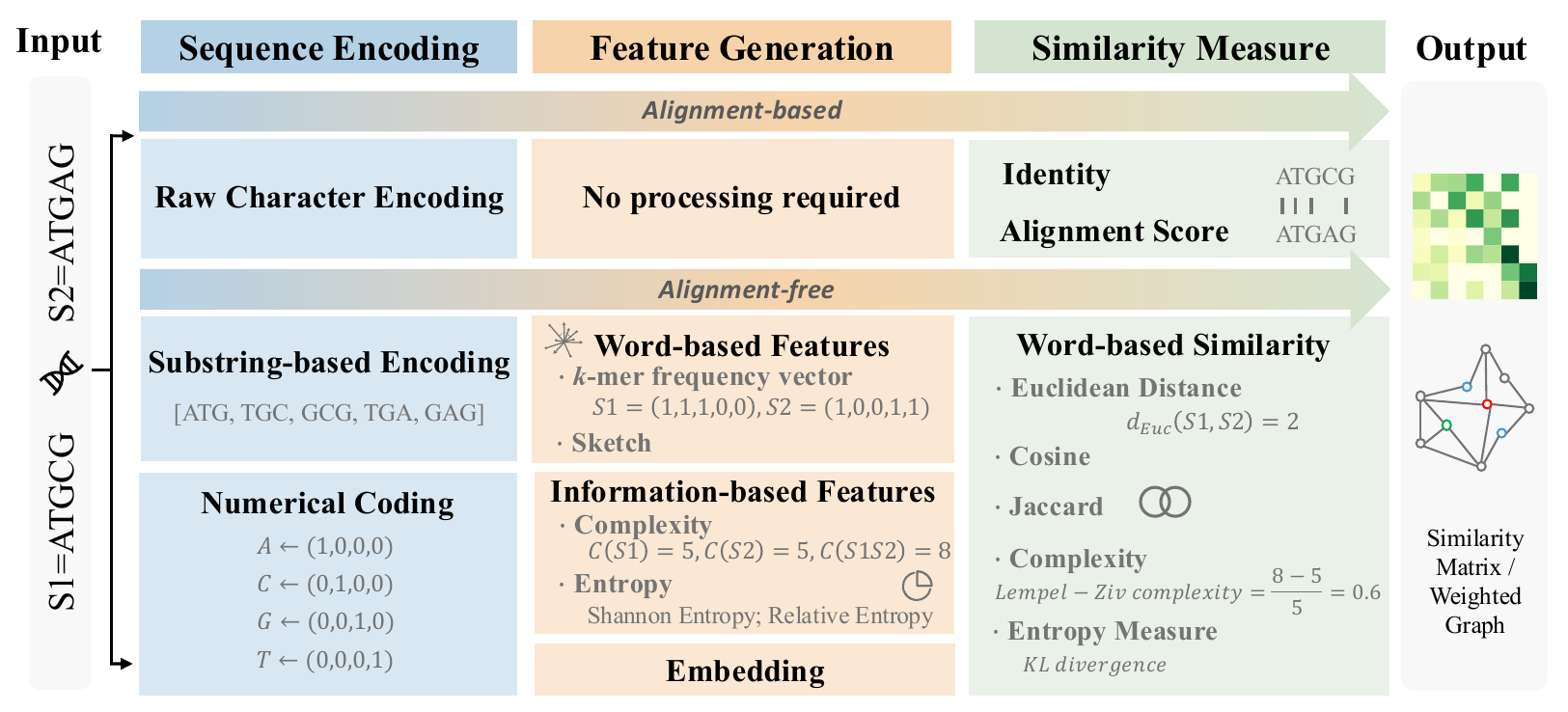} 
	\caption{Similarity modeling framework for biological sequences. The analytical pipeline is organized into three steps: sequence encoding, feature generation, and similarity measurement. Given two input DNA sequences (S1: ATGCG and S2: ATGAG), the framework branches into alignment-based and alignment-free paradigms. Alignment-based methods operate on raw character encodings and compute similarity directly through sequence identity or alignment scores. In contrast, alignment-free approaches either apply substring-based encoding to derive word-based features, such as 3-mer frequency vectors (S1=(1,1,1,0,0), S2=(1,0,0,1,1)) or sketches, or extract information-based features, for example, Shannon entropy and sequence complexity. Similarity is then quantified using a range of metrics, such as Euclidean distance, Jaccard similarity, complexity, and Kullback–Leibler divergence. The resulting similarity information is ultimately represented as a similarity matrix or a weighted graph, which serves as the input for subsequent clustering analyses.} 
	\label{simi} 
\end{figure*}

Similarity modeling is central to biological sequence clustering and represents its distinction from conventional clustering tasks. The way similarity is defined reflects different trade-offs among evolutionary principles, information preservation, and computational efficiency.
Different from existing reviews on alignment-free sequence comparison, which primarily focus on similarity computation itself, this work adopts a clustering-centered perspective and considers similarity modeling within the context of the entire clustering workflow. We emphasize that the deployment of different similarity measures can affect the choice of subsequent clustering mechanisms. Building on prior studies \cite{zielezinski2017alignment} \cite{alignfree2023}, we propose a unified framework that integrates both alignment-based and alignment-free similarity methods and clarifies how different similarity modeling choices impose structural constraints on the clustering strategies that can follow.
From a methodological perspective, as shown in Fig.~\ref{simi}, most biological sequence similarity analysis procedure can be viewed as comprising three steps: \emph{sequence encoding}, \emph{feature generation}, and \emph{similarity measurement}. 

\subsection{Sequence Encoding: From Biological Sequences to Computational Objects}
The first step in similarity modeling is to transform biological sequences into forms that can be directly processed by algorithms. This process, commonly referred to as sequence encoding, focuses on how a sequence is represented. Based on existing approaches, sequence coding can be broadly categorized into the following forms.

\begin{itemize}
    \item \textbf{Raw character encoding}: Sequences are directly treated as strings composed of a finite alphabet, such as A, T, C, and G, forming the basis for alignment-based similarity modeling.

    \item \textbf{Substring-based encoding}: To avoid the computational burden of alignment, many alignment-free methods segment sequences into fixed length contiguous substrings, such as $k$-mers \cite{blaisdell1989average} \cite{deschavanne1999genomic}, and use these substrings as basic units for analysis. In this representation, a sequence is converted into a collection of substrings, which facilitates frequency counting, statistical modeling, and approximate similarity computation. This form of encoding does not rely on global alignment and serves as the starting point for most alignment-free feature extraction methods.

    \item \textbf{Numerical encoding}: Some approaches further map symbolic sequences into numerical representations \cite{mendizabal2017dna,shi2012dna,chakravarthy2004autoregressive}, for example by assigning each nucleotide or amino acid a numeric value or vector. This type of encoding is not intended to uncover sequence structure, but rather acts as an intermediate step that provides numerical input for statistical estimation, distance calculation, or information theoretic analysis. 
\end{itemize}

\subsection{Feature Generation: What Information Should Be Preserved?}

The key question in the feature generation stage is which information should be extracted from sequences and retained for subsequent similarity computation. It is important to note that this stage does not exist as an independent step in all similarity modeling approaches.

In alignment-based approaches, similarity is computed directly through character level alignment. During the search for an optimal alignment, matching, mismatching, and gap events are implicitly evaluated and weighted by the alignment algorithm. As a result, the final alignment score or identity can be viewed as a similarity measure in itself. In this sense, alignment-based methods usually merge feature generation and similarity computation into a single process.

By contrast, in most alignment-free approaches, feature generation becomes a critical intermediate stage. When explicit character by character alignment is no longer performed, sequence properties must be captured and represented in alternative forms that enable efficient comparison. In this setting, features are designed to summarize sequence composition or statistical patterns. According to their underlying construction principles, these features can be broadly grouped into three categories \cite{vinga2003alignment} \cite{vinga2014information}: word-based features, information-based features and embedding.

\subsubsection{Word-based Features}

Word-based features represent the most widely used class of alignment-free approaches. The central idea is to decompose a sequence into fixed length substrings, typically referred to as $k$-mers, and to characterize sequences by quantifying the occurrence patterns of these substrings. Word-based features may take the form of $k$-mer frequency vectors \cite{blaisdell1989average} or a more customizable form \cite{luczak2019survey} \cite{reinert2009alignment,bonham2014alignment,wang2008wse,aita2011mathematical}.

These approaches are intuitive and computationally efficient, and they effectively capture local sequence patterns. Because they do not depend on global sequence alignment, they are generally more robust to large variations in sequence length and to structural rearrangements. Nevertheless, they exhibit inherent limitations. Increasing the value of $k$ enhances discrimination of local patterns but also causes the feature space to grow exponentially, resulting in increasingly sparse shared $k$-mer matches. Furthermore, since $k$-mers are inherently local and fixed window statistics, they have limited capacity to model long range dependencies within sequences.

When dealing with massive datasets, constructing full $k$-mer frequency vectors is often infeasible due to prohibitive memory and computational requirements. To mitigate this issue, a variety of approximation and compression strategies have been proposed. For long sequences, utilizing sparse suffix arrays can identify maximal exact matches between sequences \cite{khan2009practical}, thereby reducing memory consumption.
More recently, to support large-scale biological sequence analysis, hashing and random sampling strategies have been widely adopted to produce low-dimensional and compact sequence representations. These approaches, commonly referred to as ``sketch'' \cite{rowe2019levee}, probabilistically sample and compress the underlying $k$-mer set to obtain approximate representations of sequences. The classical MinHash algorithm is a representative example and can be viewed as an extension of the word-based feature generation paradigm.

\subsubsection{Information-based Features}
In contrast to word-based methods that rely on substring statistics, information-based features offer an alternative perspective for feature construction. Rather than counting the frequencies of specific patterns, information-based features can be generated by entropy or complexity \cite{lempel2003complexity}, capturing properties such as regularity, redundancy, and randomness.

Concepts from information theory can be used to characterize the properties of biological sequences. For example, Shannon entropy measures the average information content of nucleotide symbol distributions in a DNA sequence. Relative entropy can be used to quantify the information difference between the nucleotide frequency distribution of a given DNA sequence and the corresponding distribution over the entire dataset. In addition, from a complexity perspective, Kolmogorov complexity can be employed to measure sequence complexity by the length of its shortest description.

Because they focus on global statistical properties rather than local pattern matching, information-based features tend to be less sensitive to isolated mutations or short noisy segments and often provide more stable assessments of overall similarity. On the other hand, this global perspective may limit their ability to capture subtle differences between highly similar sequences. When fine grained structural variations are of interest, methods based on explicit word level statistics are often more intuitive and effective.

\subsubsection{Embedding}

In recent years, the widespread adoption of deep learning in bioinformatics has introduced a new paradigm for biological sequence similarity computation: representation learning \cite{karim2021deep} \cite{pereira2025large}. Methods under this paradigm typically map raw sequences into a continuous low-dimensional embedding space, implicitly encoding their statistical properties, contextual dependencies, and latent structural patterns. Beyond similarity assessment, embedding-based representations also shape downstream clustering strategies. By projecting biological sequences into a shared vector space, these methods allow classical clustering algorithms, such as partitioning-based techniques designed for numerical data, to be applied directly, without relying on pairwise sequence alignment.

\subsection{Similarity Measures: From Features to Clusterable Relationships}

Once feature representations have been obtained, the similarity measurement stage transforms these representations into explicit numerical relationships. At this point, similarity is no longer an abstract concept but is expressed in concrete quantitative forms, such as distances or similarity scores, which can be directly used for clustering.

\subsubsection{Alignment-based Similarity Measures}
As discussed above, in alignment-based approaches, similarity measures are derived directly from the results of sequence alignment. The core idea of these methods is to model the evolutionary process through character level alignment. One of the most commonly used measures is \emph{Identity}, typically defined as the proportion of matching characters in an optimal alignment. 

In addition to \emph{Identity}, many methods use \emph{alignment score} as similarity measures. Similarity score is usually computed based on substitution matrices and gap penalty schemes, and may be further normalized by sequence length to enable comparisons across sequences of different lengths. Compared with \emph{Identity} alone, \emph{alignment score} can distinguish between different types of mismatches and gap events. However, the numerical values depend strongly on the chosen scoring system and parameter settings.

Despite their strong biological interpretability, alignment-based similarity measures are computationally expensive (there are $\frac{(2N)!}{(N!)^2}$ different ways to form a gapped alignment between two sequences of length $N$) \cite{lange2002mathematical} \cite{chatzou2016multiple}. As a result, constructing complete similarity matrices using alignment-based methods is often infeasible for large scale datasets.

\subsubsection{Alignment-free Similarity Measures}

In alignment-free sequence analysis, similarity modeling is typically performed on top of feature representations generated from biological sequences. As discussed in the feature generation stage, both word-based and information-based approaches transform sequences into vector or set representations that encode compositional, statistical, or structural properties of the underlying symbolic strings. Once such representations are obtained, sequence similarity can be considered as a general comparison problem in feature space. This formulation decouples similarity computation from biological sequence alignment, allowing a wide range of generic similarity or distance measures to be applied.

For word-based features, commonly used choices include Euclidean distance and cosine similarity, which capture absolute and directional differences between feature vectors, respectively. In practice, cosine similarity is often preferred due to its reduced sensitivity to scale variation and sequence length effects. More general distance or similarity measures, such as Mahalanobis distance or Pearson correlation coefficient, can also be employed, providing flexibility in adapting similarity modeling to the requirements of different clustering or downstream analytical tasks. In cases where feature representations take the form of sets, set-based measures such as Jaccard similarity provide a natural means of quantifying shared content and play a central role in sketch-based techniques. 

For information-based features, complexity-based features quantify how much information a sequence contains by measuring its compressibility. In practice, this is often approximated using universal compression algorithms such as those based on Lempel–Ziv complexity \cite{otu2003new}, where the compressed lengths of individual sequences and their concatenation are compared to define distances such as the normalized compression distance (NCD) \cite{li2004similarity}. For entropy-based features, Kullback–Leibler divergence can be utilized to quantify differences between the distributions of two sequences.

Overall, alignment-free similarity measures operate on the principle that biologically meaningful sequence relationships can be captured through appropriately designed feature representations and assessed using general-purpose similarity functions. While these measures primarily reflect global compositional or statistical properties rather than fine-grained local alignments, they offer substantial advantages in scalability and flexibility. Besides, embedding vectors can be treated as another form of alignment-free features, which provides a new perspective for similarity measurement.

The output of similarity measurement can also be transformed into other representations. For example, a similarity matrix may be interpreted as a weighted graph, inducing an explicit graph structure, or even music \cite{paul2021clustering}. In this way, similarity measures not only link feature representations to clustering algorithms, but also determine, in formal terms, the types of clustering strategies that can be applied.

\section{Clustering Mechanisms}

In this section, clustering mechanisms address the question of how algorithms use the predefined similarity relationships to partition large collections of biological sequences into clusters.

Compared with general clustering tasks, biological sequence clustering exhibits clear preferences in the choice of clustering mechanisms, largely due to the intrinsic properties of biological sequences. Biological sequences are strings, which makes clustering strategies based on computing cluster centroids and iteratively updating means inappropriate, since the ``average'' of multiple sequences cannot be directly calculated. As a result, commonly used approaches in biological sequence clustering are often organized around representative sequences and rely on similarity thresholds to determine cluster membership, making threshold-based filtering strategies the dominant paradigm in this field.

Second, biological sequences are not independent samples. The similarities among them reflect shared evolutionary history rather than simple geometric proximity. In this setting, cluster structures are better viewed as branches of an evolutionary tree cut at a particular similarity threshold, rather than as well separated regions. This perspective explains why similarity threshold-based partitioning mechanisms that preserve hierarchical relationships have long been well aligned with biological applications.


Under the combined influence of these factors, biological sequence clustering has gradually converged on several representative mechanistic paradigms.
These include greedy incremental methods centered on representative sequences, clustering approaches that emphasize hierarchical structure, and extended methods that incorporate graph representations and statistical assumptions and so on. Understanding the underlying reasons for these mechanistic preferences provides a principled basis for systematically comparing different clustering algorithms in terms of their applicability and limitations.

\begin{table*}[!ht]
\centering
\footnotesize
\caption{Greedy Incremental Biological Sequence Clustering Methods. }
\label{table1}
\begin{adjustbox}{width=\textwidth}
\renewcommand{\arraystretch}{1.4}
\begin{tabular}{p{3cm}p{2cm}p{1.5cm}p{7cm}}
\toprule
\textbf{Method} &
\textbf{Source} &
\textbf{Code} &
\textbf{Access Link} \\
\midrule

CD-HIT & \cite{li2006cd} \cite{li2001clustering,li2002tolerating,huang2010cd,niu2010artificial,fu2012cd}&
C++&\url{http://cd-hit.org} \\

UCLUST &\cite{edgar2010search} & C++&
\url{https://drive5.com/usearch/} \\

UPARSE&\cite{edgar2013uparse}& -&
\url{https://drive5.com/uparse/}\\

VSEARSH&\cite{rognes2016vsearch}&C++&
\url{https://github.com/torognes/vsearch}\\

GramClust&\cite{russell2010grammar}&C&
\url{http://bioinfo.unl.edu/}\\

SEED&\cite{bao2011seed}& C++&
\url{http://manuals.bioinformatics.ucr.edu/home/seed}\\

DySC&\cite{zheng2012dysc}&C&
\url{http://code.google.com/p/dysc/}\\

LST-HIT&\cite{namiki2013acceleration}&C++&
\url{-}\\

Swarm &\cite{mahe2014swarm} & C++&
\url{https://github.com/torognes/swarm} \\

DBH&\cite{wei2017dbh}&C++&
\url{https://github.com/nwpu134/DBH}\\

hc-OTU&\cite{park2016hc}& -&
-\\

Linclust &\cite{steinegger2018clustering} & C++&
\url{https://github.com/soedinglab/mmseqs2} \\

DMSC &\cite{wei2019dmsc} & C++&
\url{https://github.com/NWPU-903PR/DMSC} \\

DNACLUST &\cite{ghodsi2011dnaclust} & C++&
\url{https://dnaclust.sourceforge.net} \\

MSClust & \cite{chen2013msclust} & Matlab&
\url{https://zhaocenter.org} \\

LSH &\cite{rasheed201316s} & Python &
\url{https://cs.gmu.edu/~mlbio/LSH-DIV/}\\

OTUCLUST&\cite{albanese2015micca} & Python \& C &
\url{https://compmetagen.github.io/micca/}\\

clust-greedy &\cite{rabbit} &C++&
\url{https://github.com/RabbitBio/RabbitTClust} \\

Gclust&\cite{li2019gclust}&C\&C++&
\url{https://github.com/niu-lab/gclust}\\

RAFTS$^{3}$G&\cite{de2019rafts3g}&Matlab&
\url{https://sourceforge.net/projects/rafts-g/}\\

Clusterize&\cite{wright2024accurately}&R&
\url{https://doi.org/10.18129/B9.bioc.DECIPHER}\\

nGIA &\cite{ju2021efficient} \cite{ju2022ngia}&C++&
\url{https://github.com/SIAT-HPCC/gene-sequence-clustering}\\

DIAMOND DeepClust&\cite{buchfink2023sensitive}&C++&
\url{https://github.com/bbuchfink/diamond}\\

OptiClust&\cite{westcott2017opticlust} &-&
\url{https://github.com/SchlossLab/Westcott_OptiClust_mSphere_2017}\\

\bottomrule
\end{tabular}
\end{adjustbox}
\end{table*}

\subsection{Greedy Incremental Clustering Methods}

Greedy incremental clustering represents the most widely used class of methods in biological sequence clustering.
The details of the algorithms, including implementation languages and access links, are summarized in Table \ref{table1}. 
The core idea is to process sequences sequentially and compare each incoming sequence with a set of existing representative sequences. If the similarity to any representative exceeds a predefined threshold, the sequence is assigned to the corresponding cluster; otherwise, it initiates a new cluster and becomes a new representative. It substantially reduces computational cost since the procedure typically requires only one single pass over the data. 

Early methods mainly adopted a single representative strategy, in which each cluster is characterized by one seed sequence and new sequences are compared only against the current set of seeds. CD-HIT \cite{li2006cd} and UCLUST \cite{edgar2010search} are well known examples of this paradigm. While highly efficient, this framework is sensitive to input order and initial seed selection, and its ability to capture cluster boundaries and global structure is therefore limited.

To alleviate sensitivity to the input order, subsequent studies \cite{zheng2012dysc} \cite{chen2013msclust} \cite{wei2019dmsc} proposed to allow representative sequences to be updated or multiple representatives to describe a single cluster during the clustering process to further improve the stability. These extensions generally improve clustering quality, at the cost of increased computational complexity.

In ultra large scale settings, research efforts have further shifted toward minimizing unnecessary similarity comparisons. Linclust \cite{steinegger2018clustering}, for example, pre-groups sequences using $k$-mer based hashing and restricts comparisons to candidate subsets, thereby avoiding exhaustive scans over all representatives and achieving near linear time complexity in practice. Such methods \cite{wright2024accurately} preserve the basic greedy incremental decision logic, but significantly improve scalability through more aggressive indexing and filtering strategies.

Overall, greedy incremental clustering has evolved along a trajectory from \textbf{single-representative}, to \textbf{multi-representative}, and finally to \textbf{linear complexity variants}. While well suited for large scale sequence clustering, its reliance on local similarity-based decisions imposes inherent limitations on capturing global structures.

\begin{table*}[!ht]
\centering
\footnotesize
\caption{Hierarchical Biological Sequence Clustering Methods. }
\label{table2}
\begin{adjustbox}{width=\textwidth}
\renewcommand{\arraystretch}{1.4}
\begin{tabular}{p{3cm}p{2cm}p{1.5cm}p{7cm}}
\toprule
\textbf{Method} &
\textbf{Source} &
\textbf{Code} &
\textbf{Access Link} \\
\midrule

GeneRAGE&\cite{enright2000generage}&-&
-\\

SWORDS&\cite{chaudhuri2002swords}&-&
\url{http://www.isical.ac.in/~probal}\\

DomClust&\cite{uchiyama2006hierarchical}&C&
\url{http://mbgd.genome.ad.jp/}\\

Mothur & \cite{schloss2009introducing} \cite{schloss2005introducing} & C++&
\url{https://mothur.org} \\

SYSTERS&\cite{krause2005large} & - &
\url{http://systers.molgen.mpg.de/}\\

CLUSS&\cite{kelil2007cluss}&C++ &
\url{http://prospectus.usherbrooke.ca/CLUSS}\\

RCM&\cite{albayrak2010clustering}&-&
-\\

SLP&\cite{huse2010ironing}& Perl&
\url{http://vamps.mbl.edu/resources/software.php}\\

ESPRIT&\cite{sun2009esprit}&-&
\url{https://www.acsu.buffalo.edu/%7Eyijunsun/lab/software.html}\\

ESPRIT-Tree&\cite{cai2011esprit} & - &
\url{https://www.acsu.buffalo.edu/%7Eyijunsun/lab/software.html} \\

ESPRIT-Forest&\cite{cai2017esprit}&-&
\url{https://www.acsu.buffalo.edu/%7Eyijunsun/lab/software.html}\\

SynerClust&\cite{georgescu2018synerclust}&Python&
\url{https://github.com/broadinstitute/SynerClust/wiki}\\

SLAD&\cite{zheng2019parallel}& - &
\url{https://github.com/vitmy0000/SLAD}\\

3gClust&\cite{halder20183gclust}& -&
\url{https://sites.google.com/site/bioinfoju/projects/3gclust}\\

MC-UPGMA&\cite{loewenstein2008efficient}&C++ &
-\\

mBKM&\cite{wei2012novel}&-&
-\\

kClust&\cite{hauser2013kclust}&-&
\url{ftp://toolkit.lmb.uni-muenchen.de/pub/kClust/}\\

mcClust&\cite{cole2014ribosomal}& Java&
\url{https://github.com/rdpstaff/Clustering}\\

HPC-CLUST&\cite{matias2014hpc}& C++&
\url{http://meringlab.org/software/hpc-clust/}\\

oclust&\cite{franzen2015improved}& R &
\url{https://github.com/oscar-franzen/oclust/}\\

MCSC&\cite{lafond2017new}&Perl&
\url{https://github.com/Lafond-LapalmeJ/MCSC_Decontamination}\\

Vclust&\cite{zielezinski2025ultrafast}&C++&
\url{https://afproject.org/vclust/}\\

ProteinCluster&\cite{chen2025exploring}&Python&
\url{(https://github.com/johnchen93/ProteinClusterTools}\\

\bottomrule
\end{tabular}
\end{adjustbox}
\end{table*}

\subsection{Hierarchical Clustering Methods}

Hierarchical clustering (see Table \ref{table2}) is one of the earliest clustering strategies applied in biological sequence analysis. Its core idea is to start from pairwise sequence similarities and iteratively merge sequences in a bottom-up manner, thereby constructing a hierarchical organization of relationships. 
Unlike greedy incremental methods that prioritize efficiency and depend on input order, hierarchical clustering does not rely on selecting representatives, but instead seeks to preserve the global structure of sequence similarities.

The clustering process typically starts with each sequence forming its own cluster, and the two closest clusters are iteratively merged until all sequences are grouped into a single cluster or a predefined distance threshold. Early OTU construction methods largely followed this paradigm. Representative implementations include the hierarchical clustering pipeline in Mothur \cite{schloss2009introducing}, ESPRIT \cite{sun2009esprit} and its accelerated variant ESPRIT-Tree \cite{cai2011esprit}, as well as related approaches based on exact or approximate distance matrices. These methods are insensitive to input order, yield stable clustering results, and provide an explicit hierarchical view that is valuable for downstream analysis.

The primary strength of hierarchical clustering lies in its interpretability. By fully retaining pairwise similarity relationships, it is often regarded as a structural reference or accuracy benchmark for small scale datasets. However, its reliance on a complete pairwise similarity matrix typically leads to a quadratic time complexity,  which severely limits scalability under high throughput sequencing settings. Even with accelerations based on nearest neighbor search or tree structures, this computational bottleneck is still difficult to eliminate entirely. Moreover, the chaining effect in large scale biological sequence data can cause a substantial fraction of sequences to be merged into a single oversized cluster \cite{chen2006seqoptics}.

Consequently, in current practice, hierarchical clustering serves more often as a methodological baseline rather than a general purpose solution. It offers a clustering perspective that closely aligns with evolutionary structure, but its computational cost restricts its applicability to scenarios with limited data scale, high accuracy requirements, or comparative evaluation of other clustering methods.

\begin{table*}[!ht]
\centering
\footnotesize
\caption{Graph-based Biological Sequence Clustering Methods. }
\label{table3}
\begin{adjustbox}{width=\textwidth}
\renewcommand{\arraystretch}{1.4}
\begin{tabular}{p{3cm}p{2cm}p{1.5cm}p{7cm}}
\toprule
\textbf{Method} &
\textbf{Source} &
\textbf{Code} &
\textbf{Access Link} \\
\midrule
ProtoMap&\cite{yona2000protomap}&-&
\url{http://www.protomap.cs.huji.ac.il}\\

MCL &\cite{enright2002efficient} & C&
\url{https://micans.org/mcl} \\

ProClust&\cite{pipenbacher2002proclust}&-&
\url{http://www.bioinformatik.uni-koeln.de/~proclust/download/}\\

BAG&\cite{kim2006bag}&-&
-\\

SeqGrapheR&\cite{novak2010graph}&R&
\url{http://w3lamc.umbr.cas.cz/lamc/resources.php}\\

FORCE&\cite{wittkop2007large}& Java&
\url{http://gi.cebitec.uni-bielefeld.de/comet/force/}\\

SiLiX&\cite{miele2011ultra}& C++&
\url{http://lbbe.univ-lyon1.fr/SiLiX}\\

PanOCT&\cite{fouts2012panoct}&Perl&
\url{https://sourceforge.net/projects/panoct/}\\

M-pick&\cite{wang2013m}&-&
\url{http://plaza.ufl.edu/xywang/Mpick.htm}\\

CLUSTOM&\cite{hwang2013clustom}&C &
\url{http://clustom.kribb.re.kr}\\

Roary& \cite{page2015roary} &Perl&
\url{http://sanger-pathogens.github.io/Roary}\\

CHAINS &\cite{matar2025biological} & C++ &
\url{https://github.com/johnymatar/SpCLUST-Global} \\

clust-mst &\cite{rabbit} &C++&
\url{https://github.com/RabbitBio/RabbitTClust}\\

MtHc&\cite{wei2015mthc}& - &
\url{http://compgenomics.utsa.edu/mthc/}\\

DMclust&\cite{wei2017dmclust}&-&
-\\

DBH&\cite{wei2017dbh}&C++&
\url{https://github.com/nwpu134/DBH.git}\\

CLUSTER2&\cite{dhanda2018development}&-&
\url{http://tools.iedb.org/cluster2/}\\

BFClust&\cite{surujonu2020boundary}&-&
-\\

AncestralClust&\cite{pipes2022ancestralclust}&&
\url{https://github.com/lpipes/ancestralclust}\\

ALFATClust&\cite{chiu2022clustering}& Python&
\url{https://github.com/phglab/ALFATClust}\\

Complet+&\cite{nguyen2023complet+}& Python&
\url{https://github.com/EESI/Complet-Plus}\\

\bottomrule
\end{tabular}
\end{adjustbox}
\end{table*}

\subsection{Graph-based Clustering Methods}

Graph-based clustering methods (see Table \ref{table3}) represent biological sequences as nodes and construct edges based on pairwise sequence similarity, thereby forming a sequence similarity graph. Unlike greedy incremental methods centered on representative sequences, or hierarchical clustering approaches that impose a merging order, instead, graph-based clustering is performed directly on the connectivity structure induced by sequence similarities.

\begin{figure}[!ht] 
	\centering 
	\includegraphics[width=8.8cm]{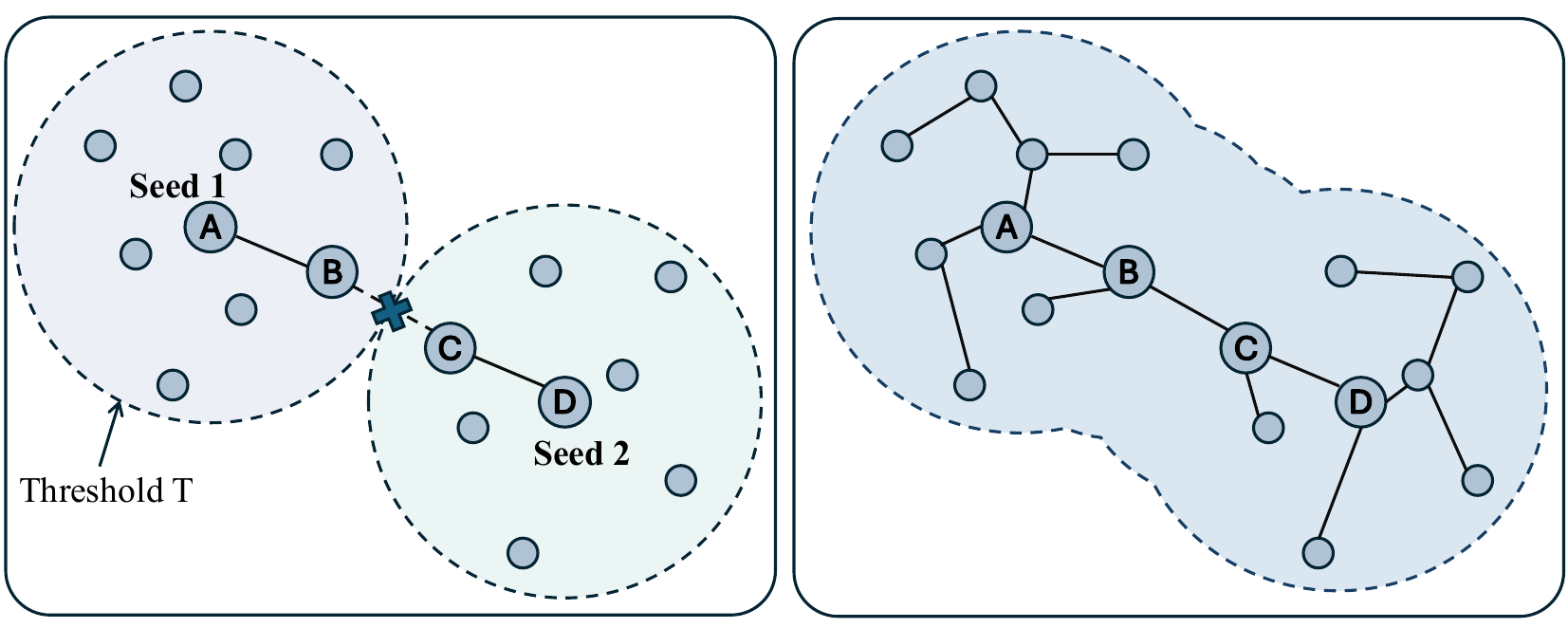} 
	\caption{Comparison between greedy incremental-based and graph-based clustering mechanisms. (Left) Greedy incremental-based strategies (e.g., CD-HIT, UCLUST) partition sequences into clusters using central seeds. This local decision-making often overlooks connections between clusters, which can fragment continuous evolutionary lineages. (Right) Graph-based methods define clusters based on overall connectivity. By capturing the connected relationship (e.g., A–B–C–D), these approaches can recover a continuous sequence relationship that connects divergent members through multiple intermediate sequences, effectively resolving evolutionary structures.} 
	\label{graph} 
\end{figure}

A notable advantage of graph-based clustering in biological sequence analysis is its ability to identify contiguous structures formed through chains of intermediate sequences. In real-world datasets, similar biological sequences often organize along evolutionary paths, giving rise to chain or network structure. In such cases shown in Fig.~\ref{graph}, sequences may be assigned to different clusters by greedy methods, even though they belong to the same contiguous sequence group when viewed from a global connectivity perspective.

By explicitly constructing a sequence similarity graph, graph-based methods preserve pairwise similarity relationships before performing clustering, and then partition the data according to the global connectivity. Clustering can be achieved based on connected components, community structure, or random walk-based criteria. Compared with strategies that assign sequences sequentially to representative seeds, these methods are better able to group sequences that are connected through chains of intermediate sequences into the same cluster.

From a methodological perspective, the key advantage of graph-based methods lies in the decoupling of similarity assessment from the final clustering decision. Rather than assigning sequences to clusters immediately upon comparison, similarity information is retained as part of a global structure and contributes to clustering at a later stage. This separation reduces misassignments and provides greater stability when dealing with complex structure.

\begin{table*}[!ht]
\centering
\footnotesize
\caption{Model-based Biological Sequence Clustering Methods. }
\label{table4}
\begin{adjustbox}{width=\textwidth}
\renewcommand{\arraystretch}{1.4}
\begin{tabular}{p{3cm}p{2cm}p{1.5cm}p{7cm}}
\toprule
\textbf{Method} &
\textbf{Source} &
\textbf{Code} &
\textbf{Access Link} \\
\midrule

CROP &\cite{hao2011clustering} & C++ &
\url{https://code.google.com/archive/p/crop-tingchenlab/} \\

BEBaC &\cite{cheng2012bayesian} & - &
\url{http://www.helsinki.fi/bsg/software/BEBaC} \\

SOHMMM&\cite{ferles2013scaled}&-&
-\\

TRIBE-MCL&\cite{szilagyi2014fast}&-&
-\\

Subspace Clustering&\cite{wallace2015application}& - &
-\\

DP-means&\cite{jiang2017dace}&C++&
\url{https://github.com/tinglab/DACE}\\

GibbsCluster&\cite{andreatta2017gibbscluster}&-&
\url{http://www.cbs.dtu.dk/services/GibbsCluster-2.0}\\

GClust &\cite{bruneau2018clustering} & Python &
\url{https://github.com/SergeMOULIN/clustering-tool-for-nucleotide-sequences-using-Laplacian-Eigenmaps-and-Gaussian-Mixture-Models} \\

SpCLUST&\cite{matar2019spclust} & Python&
\url{https://github.com/johnymatar/SpCLUST}\\

SpCLUST-V2&\cite{matar2024optimized}&C++&
\url{https://github.com/johnymatar/SpCLUST-V2/tree/master}\\
\bottomrule
\end{tabular}
\end{adjustbox}
\end{table*}

\subsection{Model-based Clustering Methods}
Model-based clustering methods (see Table \ref{table4}) assume that the dataset is generated from a mixture of underlying probability distributions. Each cluster is associated with a parametric model, and clustering is achieved through parameter estimation and model inference. Unlike the above approaches, model-based methods focus on characterizing the statistical properties of clusters, thereby reframing clustering as a problem of model selection and parameter estimation. Model-based clustering methods are advantageous when cluster boundaries overlap or when probabilistic interpretations of clustering results are desired.

However, the similarity structure among biological sequences is often highly complex and difficult to capture with simple parametric distributions, making performance strongly dependent on the validity of the generative assumptions. In addition, computational demands also limit the direct application of these methods to large-scale datasets. As a result, model-based clustering is currently more commonly applied in small or medium-scale analyses.

Overall, by offering a statistical perspective and probabilistic interpretability, model-based clustering provides a complementary and valuable viewpoint for biological sequence clustering.

\begin{table*}[!ht]
\centering
\footnotesize
\caption{Partitional Biological Sequence Clustering Methods. }
\label{table5}
\begin{adjustbox}{width=\textwidth}
\renewcommand{\arraystretch}{1.4}
\begin{tabular}{p{3cm}p{2cm}p{1.5cm}p{7cm}}
\toprule
\textbf{Method} &
\textbf{Source} &
\textbf{Code} &
\textbf{Access Link} \\
\midrule

\cite{fayech2009partitioning}&\cite{fayech2009partitioning}&Java&
-\\

GSP&\cite{mendizabal2018genomic}&Matlab&
\url{https://github.com/starsudg/STARS.git}\\

\cite{paul2021clustering}&\cite{paul2021clustering}&Matlab &
-\\

DCMR&\cite{dasari2022mapreduce}&-&
-\\

Anchor Clustering&\cite{chang2024anchor}&Python&
\url{https://github.com/skylerchang/Anchor_Clustering_Nt}\\

\bottomrule
\end{tabular}
\end{adjustbox}
\end{table*}

\subsection{Partitional Clustering Methods}

Partitional clustering, with $k$-means as a representative example, is among the most widely used approaches in the field of cluster analysis. These methods divide a dataset into a predefined number of clusters and try to optimize the partition using different search strategies.  

However, partitional clustering approaches have long played a relatively marginal role in biological sequence clustering. This is largely due to the intrinsic characteristics of biological sequence data: sequences vary in length, and similarity is typically defined through sequence alignment rather than vector-based distances. In this context, the notion of a cluster center or mean, which is central to partitional methods, does not naturally correspond to a biologically meaningful representative sequence, thereby limiting the interpretability and applicability of such methods in sequence clustering tasks. In some studies \cite{sherif2022unsupervised} \cite{bielinska202520d}, deep learning models are used to extract features, followed by partitional clustering on the learned representations, yielding encouraging clustering performance.

Several studies (see Table \ref{table5}) have investigated the incorporation of partitional methods into biological sequence clustering and have demonstrated their potential under certain conditions. Nevertheless, applying partitional approaches to sequence data presents a set of new challenges. First, these methods usually require the number of clusters to be specified in advance, whereas in biological sequence analysis the true number of families or groups is often unknown and may not admit a unique solution across different similarity scales. Second, partitional methods are sensitive to initialization and susceptible to local optima, a limitation that becomes more pronounced in the presence of substantial noise. Finally, because these methods are poorly suited to nonconvex cluster structures, they tend to be less robust in low similarity regimes or in datasets with complex evolutionary relationships. As a result, partitional approaches are better positioned as complementary tools in the biological sequence clustering framework.

\begin{table*}[!ht]
\centering
\footnotesize
\caption{Deep Learning-based Biological Sequence Clustering Methods. }
\label{table6}
\begin{adjustbox}{width=\textwidth}
\renewcommand{\arraystretch}{1.4}
\begin{tabular}{p{3cm}p{2cm}p{1.5cm}p{7cm}}
\toprule
\textbf{Method} &
\textbf{Source} &
\textbf{Code} &
\textbf{Access Link} \\
\midrule

DeLUCS&\cite{millan2022delucs}&Python&
\url{https://github.com/pmillana/DeLUCS}\\

$i$DeLUCS&\cite{millan2023delucs}&Python&
\url{https://github.com/Kari-Genomics-Lab/iDeLUCS}\\

DeepSeqProt&\cite{david2023unsupervised}&Python&
\url{https://github. com/KyleTDavid/DeepSeqProt}\\

CGRclust&\cite{alipour2024cgrclust}&Python&
\url{https://github.com/fatemehalipour/CGRclust}\\

\bottomrule
\end{tabular}
\end{adjustbox}
\end{table*}

\subsection{Deep learning-based Clustering Methods}

In recent years, deep learning-based approaches (see Table \ref{table6}) have gradually emerged as a distinct clustering paradigm \cite{pereira2025large} \cite{karim2021deep}. Such approaches are particularly valuable for challenging tasks such as identifying remotely homologous sequences, since when identity falls into the 20–35\% range, often referred to as the ``twilight zone'', true homologs become difficult to distinguish from unrelated sequences \cite{pereira2025large} \cite{zielezinski2017alignment}. By capturing complex evolutionary patterns, deep learning models can identify family members that exhibit sequence divergence yet share similar structure.

Despite these advantages, several challenges remain when applying deep learning to biological sequence clustering. First, biological sequences are discrete strings, and converting them into numerical representations \cite{sherif2022unsupervised} \cite{bielinska202520d} that preserve biological meaning remains a challenge. Second, model performance is highly sensitive to the composition, scale, and distribution of training data. Differences across datasets can easily lead to unstable clustering results and limit the transferability of learned representations to new data. In addition, training deep learning models is computationally intensive, and the resulting clustering outcomes often suffer from limited interpretability, complicating result validation and biological interpretation.

Consequently, deep learning-based methods offer new directions for scenarios involving clustering highly diverse sequences and challenging similarity modeling. However, in applications where interpretability, stability, or computational efficiency are primary concerns, traditional clustering methods continue to play an indispensable role.

\begin{table*}[!ht]
\centering
\footnotesize
\caption{Miscellaneous Clustering Methods. }
\label{table7}
\begin{adjustbox}{width=\textwidth}
\renewcommand{\arraystretch}{1.4}
\begin{tabular}{p{3cm}p{2cm}p{1.5cm}p{7cm}}
\toprule
\textbf{Method} &
\textbf{Source} &
\textbf{Code} &
\textbf{Access Link} \\
\midrule

SEQOPTICS&\cite{chen2006seqoptics}&-&
-\\

Spark-Based \\DBSCAN Algorithm&\cite{sekhar2020analysis}&-&
-\\

DPCfam&\cite{russo2021density} \cite{barone2024protein} &C++&
\url{https://gitlab.com/ETRu/dpcfam}\\

ClusterPicker&\cite{ragonnet2013automated}&Java&
\url{http://hiv.bio.ed.ac.uk/software.html}\\

TreeCluster&\cite{balaban2019treecluster}&-&
\url{https://github.com/niemasd/TreeCluster}\\

HmmUFOtu&\cite{zheng2018hmmufotu}&C++&
\url{https://github.com/Grice-Lab/HmmUFOtu/}\\

SCRAPT &\cite{luan2023scrapt} & Python &
\url{https://github.com/hsmurali/SCRAPT} \\

\cite{barash2004meanshift}&\cite{barash2004meanshift}&-&
-\\

MeShClust &\cite{james2018meshclust}\cite{girgis2022meshclust} &C++ &
\url{https://github.com/BioinformaticsToolsmith/Identity}\\

CC&\cite{rahman2017metagenome}&Python&
\url{https://github.com/mrahma23/LSH-Canopy}\\

panX&\cite{ding2018panx}&Python&
\url{pangenome.de}\\

EdtClust&\cite{xiang2023edtclust}&C++&
\url{https://github.com/CuteYisin/EdtClust}\\

\bottomrule
\end{tabular}
\end{adjustbox}
\end{table*}

\subsection{Miscellaneous Clustering Methods}

Beyond the six major types of clustering methods discussed above, many other approaches introduce additional innovations in the clustering workflow  (see Table \ref{table7}). For example, some methods adopt iterative representative update strategies, such as MeShClust \cite{girgis2022meshclust} and SCRAPT \cite{luan2023scrapt}. By dynamically updating cluster representatives or centroids, these methods allow sequences to be reassigned during the clustering process, thereby improving clustering quality. 

Some studies \cite{balaban2019treecluster} \cite{zheng2018hmmufotu} have proposed clustering strategies based on phylogenetic trees, which incorporate tree structures to more naturally reflect evolutionary distances and branching relationships among sequences, thereby improving the accuracy and consistency of biological sequence clustering. However, these methods rely heavily on the quality of the input phylogenetic tree, and inaccuracies or biases in tree inference can directly propagate to the clustering results.

More recently, density-based clustering methods such as DBSCAN \cite{ester1996density} or HDBSCAN \cite{campello2013density} are employed \cite{matar2025biological} to partition biological sequences. Similar density-based ideas have also been adopted in some applications \cite{chen2006seqoptics}. For example, AF-Cluster \cite{wayment2024predicting} applies DBSCAN during the MSA preprocessing stage to compress redundant sequences. However, when clusters in the data exhibit substantially different density levels, fixed density thresholds are often unable to accommodate all structures simultaneously, which limits the applicability of these methods to highly diverse datasets.

Some methods \cite{ding2018panx,rahman2017metagenome,xiang2023edtclust} adopt a divide-and-conquer strategy that performs coarse clustering followed by refined clustering. For example, CC \cite{rahman2017metagenome} first uses hashing to partition data into overlapping canopies, and then runs more precise clustering algorithms in parallel within each canopy. EditClust \cite{xiang2023edtclust}, in contrast, iteratively refines clustering results through multiple rounds of union–find operations.

Overall, these methods offer alternative perspectives on biological sequence clustering and introduce new ideas that may inspire further methodological development in this field.

\section{Algorithm Design Objective}

The design objective of biological sequence clustering algorithms typically revolves around three core objectives. First, an algorithm must achieve sufficient \textbf{scalability and resource efficiency} to accommodate the continuous growth of sequence data, addressing the practical question of whether the method can actually complete under realistic computational constraints. Second, the clustering process or its results should possess clear biological \textbf{interpretability}, such that the resulting clusters structures meaningfully correspond to true evolutionary or functional relationships, thereby addressing whether the clustering is biologically reasonable. Finally, the algorithm should maintain \textbf{robustness} and high clustering \textbf{quality}, effectively controlling errors and noise introduced by heuristic acceleration and approximate computations, and ensuring reliable performance across diverse data conditions. These three objectives are closely interrelated, together underpinning the practicality and effectiveness of biological sequence clustering methods.

\subsection{Scalability \& Efficiency}

In biological sequence clustering, algorithms are mainly constrained by scalability. For a dataset containing $N$ sequences, clustering approaches that rely on exact all-pair similarity computations typically incur $O(N^2)$ time and space complexity. This property does not merely imply high computational cost, it directly determines whether an algorithm is feasible in large-scale settings. Once the number of sequences exceeds a certain scale, such methods often become impractical, as they can no longer finish within realistic time and computational resource constraints. Even when individual similarity computations are highly optimized, the overall computational and memory burden still grows rapidly with data size and becomes a critical bottleneck. Consequently, the central challenge in biological sequence clustering is how to avoid or approximate exhaustive pairwise comparisons while ensuring that the clustering process itself remains tractable at scale.

To address this challenge, improving \textbf{computational efficiency} has become a primary design objective for most biological sequence clustering algorithms. The key strategy is to drastically reduce unnecessary pairwise comparisons, replacing exhaustive evaluation with heuristic or approximate procedures that trade a controlled loss in accuracy for gains in efficiency. Through such designs, algorithmic complexity can often be reduced from quadratic to linear or near-linear scale. In addition, \textbf{memory efficiency} is as critical as computational speed. The challenges arise mainly from two sources. First, the raw sequence data themselves may exceed the memory capacity of a single machine. More importantly, the intermediate data structures generated during algorithm execution, such as all-pair distance matrices (requiring $O(N^2)$ memory), $k$-mer indices, or large-scale similarity graphs, often exhibit quadratic space complexity, expanding much faster than the raw data size.
Accordingly, a wide range of methods have been proposed from different perspectives:

\begin{itemize}
    \item Greedy incremental clustering (see Table \ref{table1}) based on representative sequences has become the most widely used strategy. This approach only compares each new sequence with a set of cluster representatives, reducing the number of comparisons from $O(N^2)$ to $O(N \times R)$, where $R$ is the number of representatives and typically much smaller than $N$. 
    \item Partitional clustering methods (see Table \ref{table5}) aim to reduce comparisons through space subdivision. These methods divide the original $N$ sequences into $K$ non-overlapping subsets and iteratively optimize clusters using centroids or medoids as representatives, thereby limiting comparisons to local regions rather than performing a global search.
    \item Many methods \cite{chen2006seqoptics} adopt a coarse-to-fine, multi-stage pipeline that performs hierarchical filtering and refinement. In such designs, the first stage typically employs fast candidate-pair identification strategies, such as sketch-based \cite{rabbit} similarity estimation, to quickly eliminate clearly unrelated sequence pairs, which are then subjected to more computationally expensive and accurate comparisons. For example, Linclust \cite{steinegger2018clustering} achieves near-linear-time clustering through $k$-mer sorting and segmented matching, while Clusterize \cite{wright2024accurately} partitions sequences based on shared rare $k$-mers, followed by progressive sorting refinement and a final exact alignment step.
    \item By employing special data structures and indices, unnecessary comparisons can be greatly reduced. For example, hash-based indexing \cite{xiang2023edtclust} enables the efficient identification of candidate sets. More complex structures, such as suffix trees or suffix arrays \cite{li2019gclust}, systematically organize substring information and support fast retrieval of complex matching patterns. The essence of these strategies is to trade a one-time preprocessing or index construction cost for a substantial reduction in the number of subsequent similarity computations, thereby achieving a qualitative improvement in overall efficiency while maintaining reliable results.
    \item Streaming \cite{rabbit} and block-wise processing strategies divide data into manageable chunks that are processed sequentially, thereby avoiding the need to load the entire dataset into memory at once. Meanwhile, similarity computation and candidate filtering are naturally amenable to parallel execution. Accordingly, many studies \cite{zheng2019parallel} \cite{sekhar2020analysis} \cite{li2019gclust} \cite{yang2013large} accelerate clustering based on parallelism to make otherwise prohibitive computations practical. Notably, Zheng et al. \cite{zheng2019parallel} proposed a scalable divide-and-conquer framework that recursively partitions ultra-large datasets into smaller subsets for independent parallel clustering, followed by result merging, achieving substantial runtime reductions without sacrificing accuracy.
\end{itemize}

\subsection{Interpretability}
In contrast to approaches that primarily emphasize computational efficiency, many clustering methods place greater importance on whether the resulting clusters have clear and interpretable biological meaning. Under this objective, clustering is no longer viewed as a purely mathematical optimization issue, but rather as an approximation of real biological entities. In other words, an ideal clustering result should be interpretable \cite{hu2024interpretable, he2025significance,dong2025interpretable} as a set of sequences sharing common evolutionary origins or functional properties, rather than merely a statistical grouping formed under a particular similarity metric. This objective is of strong practical relevance in many bioinformatics applications, and even when such methods are not computationally optimal, they are still widely adopted because of their biological interpretability. Pursuing biological consistency in clustering results is a core design objective that distinguishes biological sequence clustering from general clustering methods. This notion of consistency is reflected at three main levels as follows.

\begin{itemize}
 \item Clustering results are often directly interpreted as proxies for biological functional units. For example, in microbial ecology, clusters are commonly treated as operational taxonomic units (OTUs) \cite{methodsreview}. Such applications implicitly assume that sequences within the same cluster share a common evolutionary origin or similar molecular functions. Accordingly, the algorithm must ensure that the chosen similarity measures and clustering mechanisms capture biological homogeneity, rather than mere numerical proximity.

 \item Biological consistency requires clusters to be representable. An ideal cluster should be clearly defined and succinctly characterized by one or a small number of representative sequences that capture its essential properties. Such representativeness is crucial for downstream analyses, including functional annotation and phylogenetic reconstruction. Many methods, including CD-HIT \cite{li2001clustering} and Linclust \cite{steinegger2018clustering}, adopt representative-sequence–based incremental clustering precisely in response to this requirement.

 \item Some approaches aim to explicitly model or preserve evolutionary hierarchies within the clustering process. Methods such as TreeCluster \cite{balaban2019treecluster} and ClusterPicker \cite{ragonnet2013automated}, which are based on phylogenetic trees, or AncestralClust \cite{chiu2022clustering}, which incorporates ancestral sequence inference, go beyond producing flat partitions. Instead, they construct hierarchical cluster structures allowing different clustering levels to reflect distinct evolutionary divergence scales. Such representations provide a more interpretable structural view for understanding biological processes, including gene family expansion and species diversification. By tightly coupling the clustering process with evolutionary models, these methods enhance the biological interpretability of clustering results at the structural level. At the same time, one issue that needs attention is how to intuitively interpret biological clustering results when the number of clusters is too large. When using tree structures or rules to interpret the results, the depth of the tree is often too deep, making it more difficult to understand the underlying reasons for the clustering outcomes \cite{zhang2026global}.
\end{itemize}

A wide range of clustering methods and analytical frameworks have been developed with a strong emphasis on biological interpretability. Hierarchical clustering approaches (summarized in Table \ref{table2}), which can capture multi-level relationships among sequences, play a crucial role in biological sequence analysis. Owing to the clear interpretability of their results, these methods are widely used in microbial community analysis and phylogenetic studies, although they often incur substantial computational costs when applied to large-scale datasets.


\subsection{Robustness \& Classification Quality}

The quality of clustering outcomes is commonly assessed using two complementary criteria: homogeneity and completeness. Homogeneity reflects cluster purity by measuring whether sequences within a cluster originate from a single true class, whereas completeness evaluates whether all members of a true class are grouped into the same cluster. Together, these metrics characterize the trade-off between avoiding incorrect merges and preventing over-fragmentation, and are widely used to assess the reliability and stability of biological sequence clustering results.

To handle the explosive growth in data volume, many clustering methods rely on heuristic acceleration strategies that introduce approximations, which can in turn affect biological accuracy and the robustness of the results.
For example, $k$-mer filtering or sketch-based methods can substantially reduce computational complexity, but their reliance on random sampling or a fixed window may cause distantly related homologous sequences to be missed due to insufficient shared $k$-mers, leading to false negatives. Similarly, representative-based greedy clustering algorithms are sensitive to input order and initial representative selection, which may yield unstable clusters and result in erroneous merging of non-homologous sequences or fragmentation of true homologs \cite{nguyen2023complet+}. 

Such errors can propagate to downstream biological analyses. In functional annotation, missing remote homologs may lead to incorrect inference of gene function, while in phylogenetic analysis or species delimitation, inaccurate cluster assignments can distort the reconstruction of evolutionary relationships. Consequently, a central challenge in algorithm design is how to reduce the uncertainty introduced by acceleration strategies while fulfilling the task within acceptable computational resource constraints.

To address the above challenges, researchers have struggled from different perspectives in order to improve algorithmic robustness. Among these efforts, improving similarity modeling is one of the most direct approaches to achieving this goal. Traditional similarity definitions based on sequence identity or simple distance thresholds often fail under low sequence similarity or high-noise conditions. To address this limitation, many methods \cite{matar2019spclust} incorporate statistical or distributional features, or adopt more expressive distance or similarity functions \cite{xiang2023edtclust}, thereby improving the ability of similarity measures to distinguish sequences in complex settings and reducing errors arising from overly coarse metrics. 

Beyond similarity modeling, numerous methods enhance robustness through iterative or multi-stage strategies. A common practice is to adopt a ``coarse-to-fine'' \cite{chen2006seqoptics} pipeline, in which fast but coarse heuristic methods are first used to rapidly narrow the candidate search space, followed by more refined similarity evaluation or clustering strategies applied to much smaller subsets, enabling algorithms to produce higher-quality clustering results within acceptable computational time.

In addition, some approaches improve robustness by integrating auxiliary biological or contextual information. For example, structural features \cite{miladi2019graphclust2} or evolutionary signals can also be incorporated into the analysis. Such additional features provide extra constraints on clustering decisions when sequence similarity alone is insufficient, thereby reducing the impact of errors on the final results.

Even after an initial clustering step, there is often room to further improve the cluster structure and its biological consistency. Many existing methods, such as Linclust \cite{steinegger2018clustering} and CD-HIT \cite{li2001clustering}, tend to adopt conservative clustering strategies, resulting in high intra-cluster homogeneity but low completeness, and often producing a large number of singleton or very small clusters \cite{wright2024accurately}. To address this issue, some approaches introduce post-clustering refinement strategies that aim to adjust and correct the initial results to improve overall clustering quality. A representative example is Complet+ \cite{nguyen2023complet+}, which builds upon the existing cluster structure and searches for sequences that are highly similar to current cluster members but were filtered out or missed during the initial pipeline. This post-processing strategy allows clusters to capture related sequences more comprehensively, thereby improving sequence recall and enhancing overall cluster completeness.

\subsection{Short Summary}

Overall, the design objectives of biological sequence clustering methods depend strongly on the specific data scale, sequence types, and downstream analysis tasks. Efficient heuristic algorithms are well-suited for large-scale datasets, whereas methods that emphasize biological consistency are more appropriate for tasks such as homology analysis. At the same time, controlling and correcting the errors introduced by heuristic approximations has gradually become a critical issue in algorithm design. Clarifying the goals of different methods and their applicable scenarios facilitates an appropriate balance between efficiency and accuracy in practical applications, and provides guidance for the continued improvement of clustering algorithms under increasingly complex data conditions.

\section{Discussions and Outlooks}
In summary, biological sequence clustering methods exhibit a highly diverse methodological landscape. This diversity reflects the need to balance data scale, biological interpretability, and computational constraints. Differences in similarity modeling, clustering mechanisms, and design objectives essentially indicate how individual algorithms prioritize efficiency, interpretability, and accuracy. For example, heuristic methods based on similarity thresholds trade some precision for substantial gains in computational efficiency, making them feasible for large-scale redundancy reduction. In contrast, methods that emphasize biological consistency, although often more computationally demanding, remain crucial for tasks such as homology analysis and functional annotation. Accordingly, existing approaches are better understood as complementary methods designed for different application domains, rather than as competing alternatives.

Although biological sequence clustering methods have made substantial progress over the past two decades, several key challenges still remain unresolved in the face of rapidly growing data volumes and increasingly complex application demands.

First, scalability continues to be a major bottleneck in practice. Even highly optimized heuristic methods can achieve good efficiency through aggressive filtering and approximation, these approaches still encounter significant computational and resource constraints as data scale further increases. Moreover, most existing methods assume a static dataset and provide limited support for dynamic or stream data. As a result, newly generated sequences cannot be efficiently and reliably integrated into existing clustering results, which poses a particular challenge for continuously updated databases and real-time analysis scenarios.

Second, the reliability of clustering results remains limited. Most existing tools report only the final cluster assignments, without providing confidence levels in their outputs, making it difficult for users to assess the reliability of the clustering results. At the same time, many widely used methods, such as CD-HIT and UCLUST, are sensitive to parameter choices and the order of input sequences, which limits their robustness and can lead to unstable clustering results.

Finally, clustering performance remains limited at low sequence similarity levels. Many traditional methods rely primarily on sequence identity as the main similarity criterion, and their ability to distinguish homologous from non-homologous relationships degrades substantially as similarity decreases. This often leads to highly unbalanced cluster structures, with a few oversized clusters and many small ones. Such imbalance also diminishes downstream divide-and-conquer analyses. Consequently, developing clustering strategies that remain stable and scalable in highly divergent, low-similarity sequence spaces remains a key direction for future methodological advances.

Moreover, as similarity modeling has expanded from alignment-based techniques to statistical features and representation learning, the relationship between clustering algorithms and traditional biological concepts has also evolved. While this trend broadens the applicability of clustering methods, it raises higher demands on interpretability and clustering quality. How to maintain interpretability while incorporating increasingly complex models has therefore become a central and ongoing challenge in method development.

\section*{Acknowledgments}
This work has been supported by the Natural Science Foundation of China under Grant No. 62472064.

%

\bibliographystyle{IEEEtran}
\bibliography{IEEEabrv,ref}


\vfill

\end{document}